\documentclass[lettersize,journal]{IEEEtran}
\usepackage{amsmath,amsfonts}
\usepackage{mathrsfs}
\usepackage{algorithmic}
\usepackage{algorithm}
\usepackage{array}
\usepackage[caption=false,font=normalsize,labelfont=sf,textfont=sf]{subfig}
\usepackage{textcomp}
\usepackage{stfloats}
\usepackage{url}
\usepackage{verbatim}
\usepackage{graphicx}
\usepackage{cite}
\usepackage{float}
\usepackage{datetime}
\usepackage{upgreek}
\newdateformat{monthyeardate}{\monthname[\THEMONTH] \THEYEAR}
\usepackage{hyperref}

\begin{document}

\title{Terahertz amplification by injection locking of waveguide resonant tunneling diode}

\author{James Greenberg,~\IEEEmembership{Member,~IEEE,}
Brendan M. Heffernan,~\IEEEmembership{Member,~IEEE,}
William F. McGrew,~\IEEEmembership{Member,~IEEE,}
Antoine Rolland,~\IEEEmembership{Member,~IEEE}

\thanks{The authors are with IMRA America, Inc. Boulder Research Labs 1551 S. Sunset Street, Suite C Longmont, CO, 80501 USA (e-mail: jgreenbe@imra.com).}
}

\markboth{IEEE JOURNAL OF QUANTUM ELECTRONICS,~Vol.~XX, No.~XX, \monthyeardate\today}%
{Shell \MakeLowercase{\textit{et al.}}: A Sample Article Using IEEEtran.cls for IEEE Journals}

\maketitle

\begin{abstract}
High power and low phase noise oscillators at terahertz frequencies are required for several burgeoning scientific and technological applications, including radioastronomy, imaging, molecular spectroscopy, radar, and wireless communications. Operating at terahertz oscillation frequencies presents unique challenges based on the method of generation. Electronic oscillators can produce ample power but suffer from relatively high phase noise due to the nonlinear multiplication of microwave sources. Meanwhile, photomixing of optical sources provides superior spectral purity but low usable power, due to the limited bandwidth of the photomixer. We propose a hybrid solution involving injection locking of an electronic oscillator, a resonant tunneling diode, by a low phase noise photomixed source, a dual-wavelength Brillouin laser. In this study, we demonstrate a proof-of-concept injection-locking amplifier at 260 GHz, achieving up to 40 dB gain for nanowatt-level input signals. For the first time, we characterize the residual phase noise of an injection-locked waveguide RTD, showing quantitative consistency with theoretical predictions based on detailed analysis of its free-running noise. This architecture has the potential to scale to frequencies of 1\,THz and beyond, which would provide a clear path to realize a terahertz oscillator with high power and low phase noise. 
\end{abstract}


\bstctlcite{IEEEexample:BSTcontrol}

\section{Introduction}
The realization of low-phase noise terahertz oscillators is needed for a plethora of technical and scientific benefits. Such oscillators are employed for very long baseline interferometry, such as the Event Horizon Telescope Collaboration \cite{collaboration2019}. Increasing the observation frequency above the current 345\,GHz will improve the resolution available to image supermassive black holes and other stellar objects \cite{raymond2024}. A more terrestrial application along the same lines is terahertz radar, used, for instance, to image foreign objects and debris on aircraft runways \cite{futatsumori2022}. Rotational spectroscopy of small gas-phase molecules also requires low-noise terahertz sources \cite{townes1975}. Such precision spectroscopic experiments can be leveraged into building terahertz frequency references and clocks for long-term frequency stability \cite{greenberg2025,wang2018}. Finally, wireless communications will continue to extend to higher frequency bands as more devices and higher bandwidth are needed \cite{keysight,dang2020}. 5G wireless networks and beyond demand low-noise oscillators to encode more information and optimize the signal-to-noise ratio of wireless receivers \cite{kurner2021,maekawa2024,kurner2024}.

An ideal candidate oscillator for such applications is a dual wavelength Brillouin laser (DWBL), exhibiting best-in-class phase noise from 300\,GHz to 3\,THz \cite{heffernan2024}. However, the output is an optically carried wave that requires photomixing to generate terahertz radiation. Uni-traveling-carrier photodiodes (UTCs) have recently increased their power output at 300\,GHz to be above 1\,mW by utilizing a silicon carbide substrate \cite{che2024}. Plasmonic photomixers have also realized near-mW power at 1.28\,THz \cite{huang2017}, though this is achieved at a duty cycle of a few percent, averaging 10\,$\upmu$W output, with significant spectral broadening from the photomixer's impulse response\cite{berry2014}. Frequencies at or above 1\,THz tend to show much lower output power, dropping to $\approx$10\,nW at 3\,THz. Additionally, amplifiers above the WR-3.4 band are not yet commercially available \cite{cheron2022}.

To address the power limitations of photomixing, compact electronic oscillators such as resonant tunneling diodes (RTDs) offer a promising alternative. They have demonstrated oscillation frequencies above 1\,THz with as much as 1\,mW of output power \cite{suzuki2024}. This power can also be dramatically improved by structuring RTDs into mutually self-injected arrays with over 10\,mW of coherent output power measured at 450\,GHz \cite{koyama2022}. While RTDs leave much to be desired in the way of spectral purity, as discussed in this work, the intrinsically low quality factor (Q) is favorable for injection locking. Injecting an RTD with a low-noise photomixed source has been shown to considerably improve the linewidth of an RTD \cite{hiraoka2021}. Some of the dynamics of the injection locking process have also been studied, and shown to follow the same physics as expected from microwave oscillator injection \cite{suzuki2022}. These prior studies relied on free-space setups for injection locking, severely limiting the usable power of both the photomixed source and the RTD.

In this study, we demonstrate injection locking of a waveguide RTD using a photomixed DWBL, achieving over 40\,dB amplification of nanowatt-level input signals enabled by low-loss waveguide components. We also explore the limitations of injection amplification in terms of a residual phase noise floor. We compare these measurements to theoretical predictions which rely heavily on the free-running phase noise of the RTD. A thorough analysis of the free-running RTD is also reported. While these demonstrations are at 260\,GHz, we discuss the future outlook of scaling to higher frequencies. Thus, injection amplification of photomixed sources with waveguide RTDs presents a clear path to high power and low phase noise oscillators at terahertz frequencies.
    
\section{RTD Phase noise}
\begin{figure}[t]
    \centering
    \includegraphics[width=\linewidth]{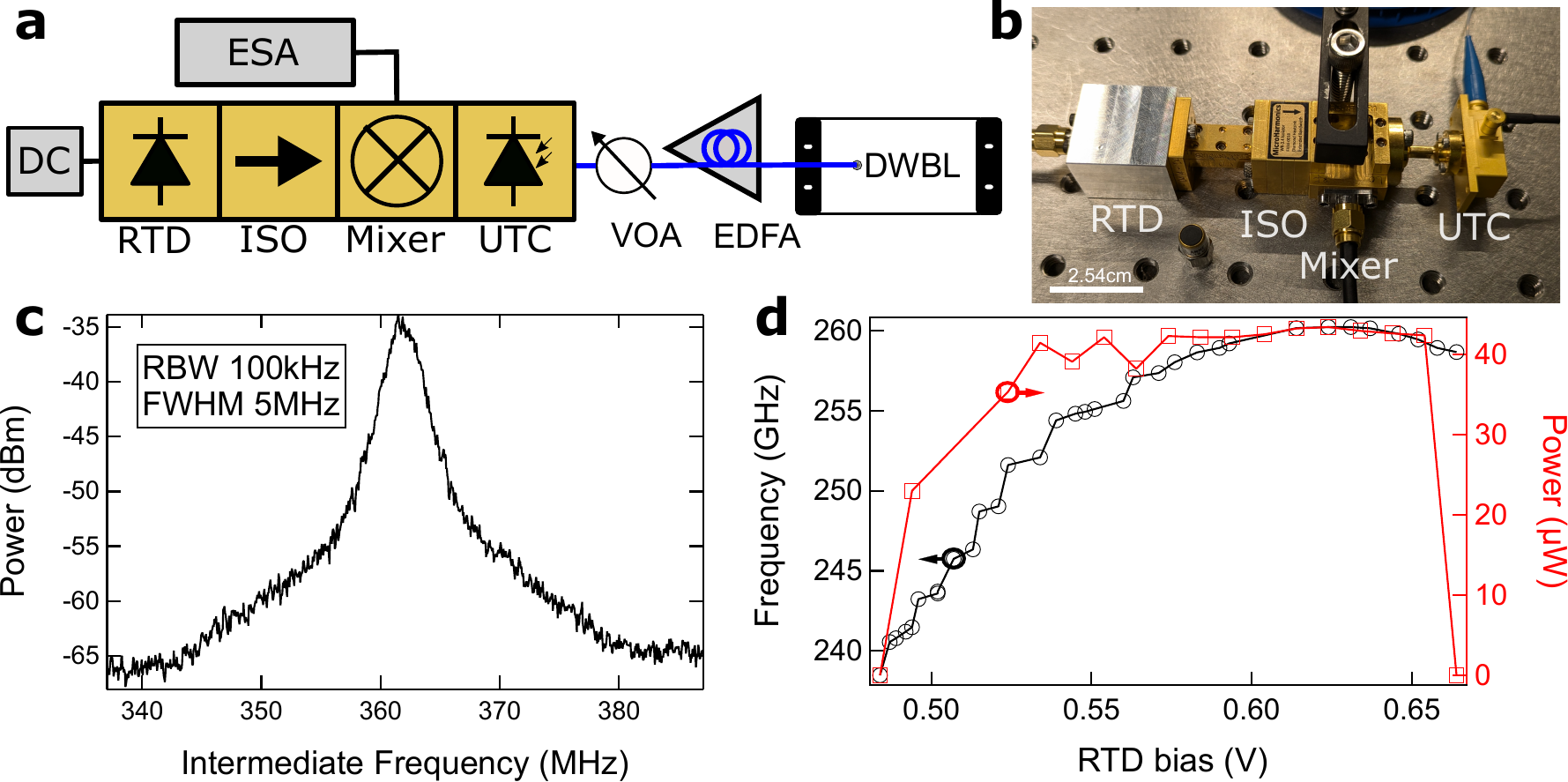}
    \caption{RTD characterization setup. \textbf{a} Schematic of the experiment, described in detail in the text. ISO is a terahertz wave isolator, VOA is a variable optical attenuator, ESA is an electrical spectrum analyzer, and EDFA is an erbium-doped fiber amplifier. \textbf{b} Photograph of the waveguide RTD and other components. \textbf{c} Electrical spectrum of the intermediate frequency full width at half maximum (FWHM) of the RTD radiation produced with a 530\,mV bias. This spectrum is the average of 100 consecutive traces. \textbf{d} RTD radiation frequency and power (left and right axes, respectively) versus applied bias voltage.}
    \label{fig_1}
\end{figure}

We first measured the free running phase noise of the RTD, enabling characterization of the residual phase noise from injection locking. The properties of the terahertz radiation were transferred to an RF intermediate frequency in the setup shown in Fig. \ref{fig_1}\textbf{a}. Radiation from a waveguide RTD (manufactured by ROHM Semiconductor) passed through an isolator (Micro Harmonics FR34M2) incident on the RF port of a fundamental mixer (VDI WR3.4BAM-ULP). The LO port was illuminated by a DWBL tuned to 250\,GHz, photomixed using a UTC (NTT IOD-PMJ-13001). The intermediate frequency, $f_\text{IF} = |f_\text{RTD} - f_\text{DWBL}|$ was measured by an electrical spectrum analyzer (Agilent N151A). Fig. \ref{fig_1}\textbf{b} is a photograph of the waveguide components, showing their geometry and compact nature. All these elements were WR-3.4 waveguide components. A typical spectrum is shown in Fig. \ref{fig_1}\textbf{c}, revealing the full width at half maximum (FWHM) of the radiation to be 5\,MHz. Fig. \ref{fig_1}\textbf{d} shows the output frequency of the RTD and power as a function of bias voltage. The operating frequency of this waveguide RTD can be tuned by over 20\,GHz. Using a 620\,mV bias, we measured a peak power of 40\,$\upmu$W at 260\,GHz. This bias was used for all subsequent measurements. The frequency data also show alternating sections of shallow and steep slopes. The steep slopes correspond to frequency jumps observed by Ref. \cite{suzuki2022} and are likely the result of frequency feedback from reflections into the RTD. The shallow slopes correspond to regions of continuous frequency tuning, with the largest region between 255 and 260\,GHz. The waveguide isolator was critical to achieving a consistent, reproducible frequency versus bias curve which otherwise can change with overall waveguide length and components.

\begin{figure}[t]
    \centering
    \includegraphics[width=\linewidth]{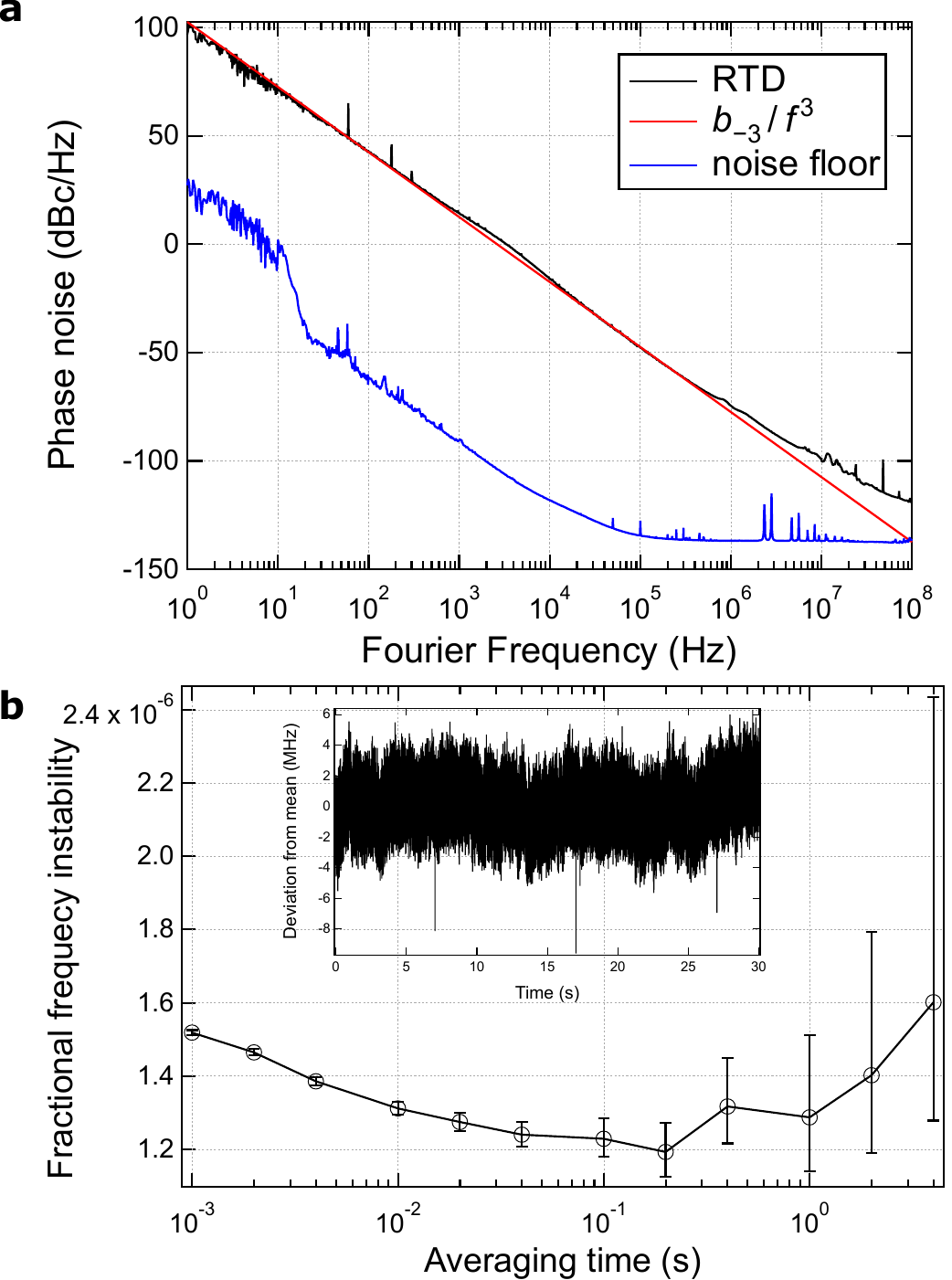}
    \caption{RTD free running phase noise at 260\,GHz and Allan deviation. \textbf{a} RTD free running phase noise and $b_{-3} = 102.5$\,dBc/Hz slope with 6 decades of quantitative agreement. See text for discussion of $b_{n}$ units. The phase noise floor of the measurement technique, arising from the free running phase noise of the DWBL and mixer, is shown below for reference. \textbf{b} Free running frequency fluctuations over 30\,s are shown in the inset. The fractional frequency instability is calculated by the Allan deviation of the free running frequency fluctuations, divided by the mean frequency (260\,GHz).}
    \label{fig_2}
\end{figure}

The free-running RTD phase noise power spectral density (PSD) was measured using a phase noise analyzer (Keysight SSA-X) in place of the electrical spectrum analyzer (ESA). The phase noise of the intermediate frequency (IF) carries that of the RTD radiation and can be expressed as
\begin{equation}
S_{\varphi}^\text{IF} = S_{\varphi}^\text{RTD} + S_{\varphi}^\text{DWBL} + S_{\varphi}^\text{mixer},
\end{equation}
where $S_{\varphi}^\text{DWBL} + S_{\varphi}^\text{mixer}$ represents the one-sided PSD of the measurement system noise, in rad$^2$/Hz. The measurement system noise floor was determined by substituting the RTD with another DWBL, as shown in Fig.~\ref{fig_2}\textbf{a} along with the measured $S_{\varphi}^\text{IF}$. The phase noise PSD was parameterized using power-law coefficients~\cite{rubiola2023}.

\begin{equation}
    S_{\varphi}(f) = \sum_{n} b_{n}f^n,
    \label{pnmodel}
\end{equation}

\noindent where $f$ is the Fourier frequency with respect to the carrier where phase noise in dBc/Hz is given by $\mathscr{L} = 10\log_{10}\left[\frac{1}{2}S_{\varphi}\right]$. The $b_n$ coefficients have units of rad$^2$Hz$^n$/Hz. For brevity and consistency with the measured data, we report the coefficients as contributions to $S_{\varphi}$ evaluated at the Fourier frequency of 1\,Hz and converted to dBc/Hz.

We found that the phase noise of the RTD follows an $f^{-3}$ trend, characteristic of flicker frequency noise, across nearly six decades of Fourier frequency. The small increase between 1\,kHz and 10\,kHz was caused by voltage noise in the DC bias supply that converts directly into frequency noise. Above 1\,MHz, there is a significant deviation into an $f^{-2}$ trend that we will discuss in the context of the Leeson model below.

To determine the stability of the RTD output frequency, the phase noise analyzer was replaced with a frequency counter. The frequency deviations relative to the mean (260\,GHz) are shown in the inset of Fig. \ref{fig_2}\textbf{b} along with the Allan deviation of those fluctuations. The relatively flat trend in the Allan deviation qualitatively matches the expectation of flicker frequency fluctuations observed in the phase noise \cite{rubiola2009}. The quantitative agreement between the phase noise PSD and allan deviation flicker floor is within 3\,dB, showing a minimum fractional instability of $1.2\times10^{-6}$, corresponding to a 310\,kHz frequency instability. At large averaging times, frequency drift was apparent, and this drift is attributed to the voltage bias drift and not the intrinsic oscillator behavior. The DWBL exhibited lower instability than the RTD at all timescales \cite{heffernan2024}, and thus did not affect the measurement.

\subsection{Leeson model}
\label{sec_leeson}
\begin{figure}[t]
    \centering
    \includegraphics[width=\linewidth]{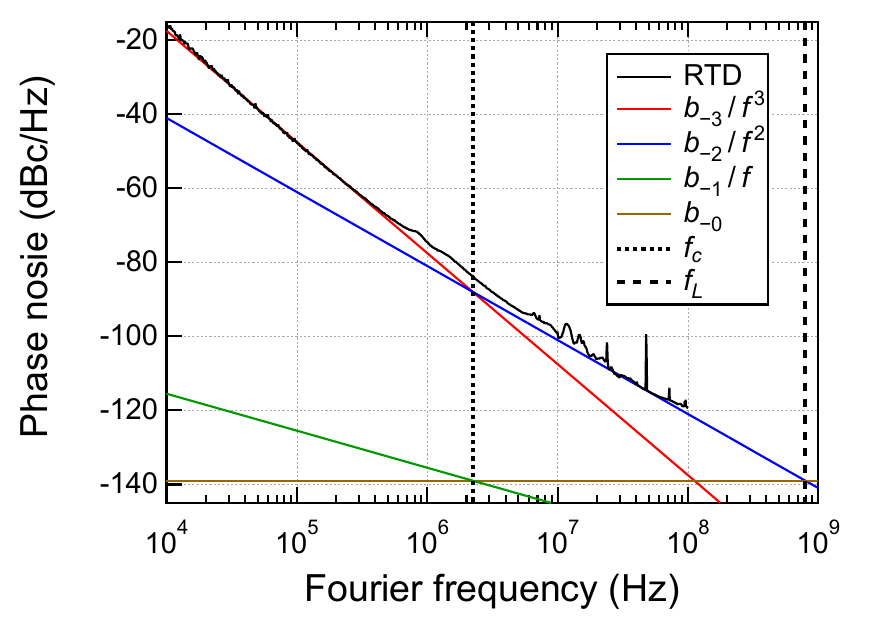}
    \caption{RTD phase noise Leeson model. The RTD free running phase noise is parameterized by $b_n$ coefficients given in dBc/Hz as explained in the text: $b_{-3} = 102.5$\,dBc/Hz and $b_{-2} = 39$\,dBc/Hz. The crossover frequency is determined to be $f_c =2.25$\,MHz and the Leeson frequency is extrapolated to be $f_L = 789$\,MHz. The Leeson effect describes the underlying diode $b_{-1} = -75.5$\,dBc/Hz noise and $b_0 = -139$\,dBc/Hz shot noise floor (also shown).}
    \label{fig_3}
\end{figure}

The waveguide RTD phase noise is described by the Leeson effect \cite{rubiola2009}, with the free-running spectrum well fit by Eq.~\ref{pnmodel} using two non-zero coefficients: $b_{-3} = 102.5$\,dBc/Hz and $b_{-2} = 39$\,dBc/Hz.  The crossover point between these two slopes is at $f_\text{c} = 2.25$\,MHz, and the intersection point is 3\,dB below the measured phase noise, as expected by the model. The relatively high crossover frequency suggests that the cavity has a low Q. While we were unable to directly detect the RTD white noise floor, due to limited ESA bandwidth and measurement noise floor, we verified by measurements of our injection locking range (see Section \ref{rtdamp}) that $Q=165$, and thus the Leeson frequency is $f_\text{L} = f_\text{RTD}/2Q =  789$\,MHz.

Underpinning this analysis is the Leeson effect,
\begin{equation}
    S_{\varphi}(f) = \left[1+\frac{f_\text{L}^2}{f^2}\right]S_{\psi}(f)
    \label{leeson}
\end{equation}

\noindent where $S_{\psi}(f)$ is the phase noise PSD of the active element inside the RTD. In this device, the diode itself provides the effective amplification through its negative resistance. Intrinsic $b_{-1}/f$ and white noise of the diode transforms into the measured, radiated phase noise via Eq. \ref{leeson}. We have included these $b_{-1}/f$ and $b_{0}$ lines in Fig. \ref{fig_3} in a self-consistent way. Direct measurement of the internal diode is not possible without disassembling the waveguide package, but the electrical characteristics should be predictive of the resulting radiation. This model suggests that the most important attributes for reducing the RTD linewidth is the diode $1/f$ noise and the $Q$ of the RTD. The white noise floor, $b_{0}$, is consistent with that of a typical diode shot noise floor with a modest enhancement factor from the resonant tunneling structure \cite{iannaccone1998}. 

\section{RTD injection and amplification}
\label{rtdamp}
\begin{figure}[t]
    \centering
    \includegraphics[width=\linewidth]{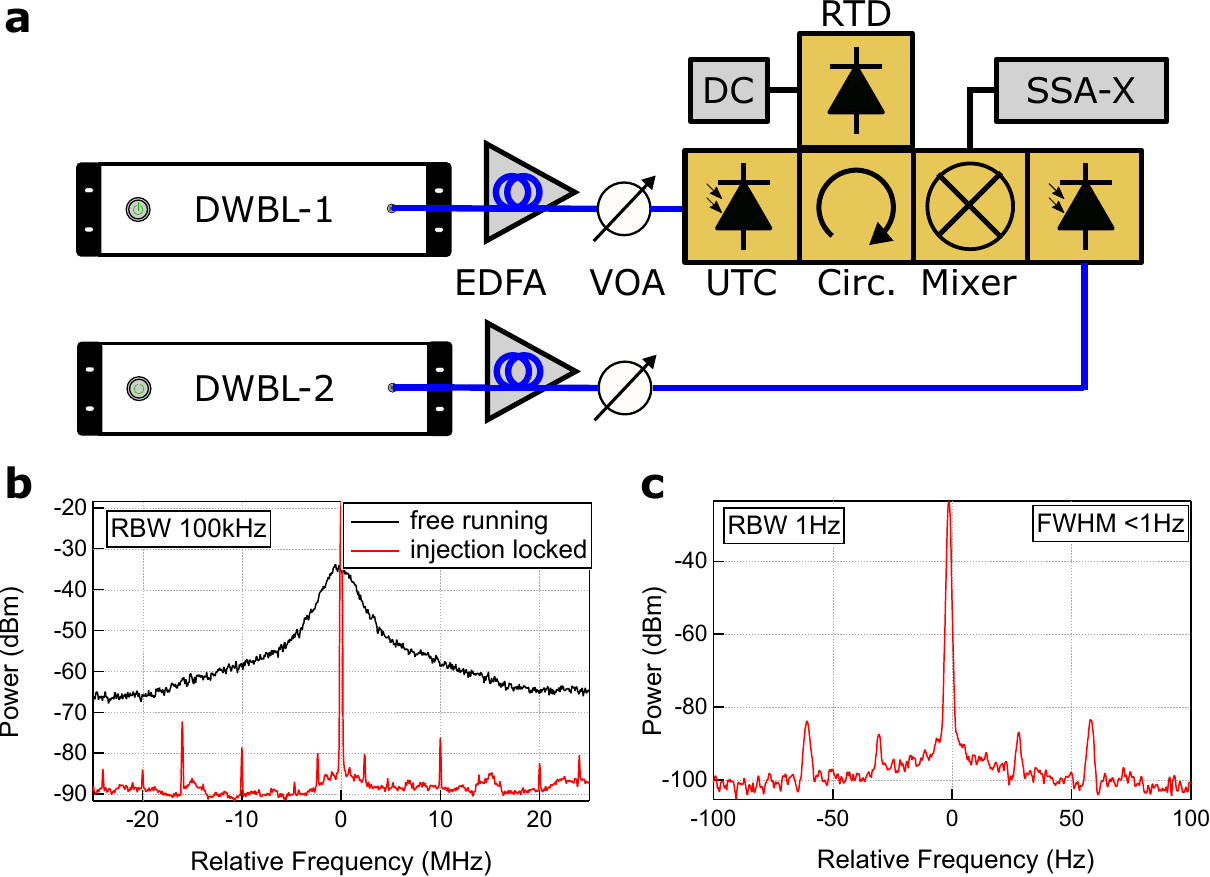}
    \caption{RTD injection locking setup and validation. \textbf{a} A schematic of the injection locking amplification setup is shown and described in detail in the text. To measure the phase noise of the injection locked RTD, a second DWBL was used as a reference. SSA-X is a phase noise analyzer. \textbf{b} ESA traces of the IF from the mixer with 100\,kHz resolution bandwidth (RBW). This shows significant reduction in FWHM as a result of injection locking. \textbf{c} A trace of the injection locked spectrum near the carrier. The determination of the FWHM was limited by the 1-Hz minimum RBW of the ESA.}
    \label{fig_4}
\end{figure}

Injection locking can be realized in a wide array of disparate physical systems, notably microwave oscillators and lasers \cite{kurokawa1968,adler1973,liu2020a}. We follow the formalism presented by Ref. \cite{chang1997a}, describing the steady state injection locking phase, given by,

\begin{equation}
    \sin{\theta} = \frac{f_\text{DWBL} - f_\text{RTD}}{f_\text{3dB}\rho},
    \label{injphase}
\end{equation}

\noindent where $f_\text{3dB} = \frac{f_\text{RTD}}{2Q}$, $Q$ is the quality factor, and $\rho=\sqrt{\frac{P_\text{DWBL}}{P_\text{RTD}}}$. In this experiment, $f_\text{RTD} = 260$\,GHz and $P_\text{RTD} = 40\,\upmu$W. These values correspond to the RTD bias voltage of 620\,mV shown in Fig. \ref{fig_1}\textbf{d}.

We used the setup shown in Fig. \ref{fig_4}\textbf{a} to injection lock the waveguide RTD to a photomixed DWBL, thereby operating as a waveguide amplifier. To minimize the injection locking phase of Eq. \ref{injphase}, a DWBL, denoted DWBL-1, was tuned to 260\,GHz and delivered to the RTD via a waveguide circulator with $<2$\,dB loss (Micro Harmonics HC034). The synchronized RTD radiation then left the circulator for the RF port of a fundamental mixer. The LO port of the mixer was driven by another DWBL, denoted DWBL-2, tuned to 250\,GHz. The 10-GHz IF from the mixer was analyzed by a phase noise analyzer (Keysight SSA-X). The IF carried the phase noise PSD of the injected RTD  $S_{\varphi}^\text{IF} = S_{\varphi}^\text{sync} + S_{\varphi}^\text{DWBL-2} + S_{\varphi}^\text{mixer}$, where $S_{\varphi}^\text{sync}$ arises from the combined phase noise of DWBL-1 and the residual phase noise of injection locking. Electrical spectra of the IF with and without injection are shown in Fig. \ref{fig_4}\textbf{b} and \textbf{c}. The FWHM was reduced from 5\,MHz to less than 1\,Hz. Injection power was varied by tuning the optical power delivered to the UTC by DWBL-1. The injection power was measured by replacing the RTD with a calibrated Schottky barrier diode (VDI ZBD-F).

\begin{figure}[t]
    \centering
    \includegraphics[width=\linewidth]{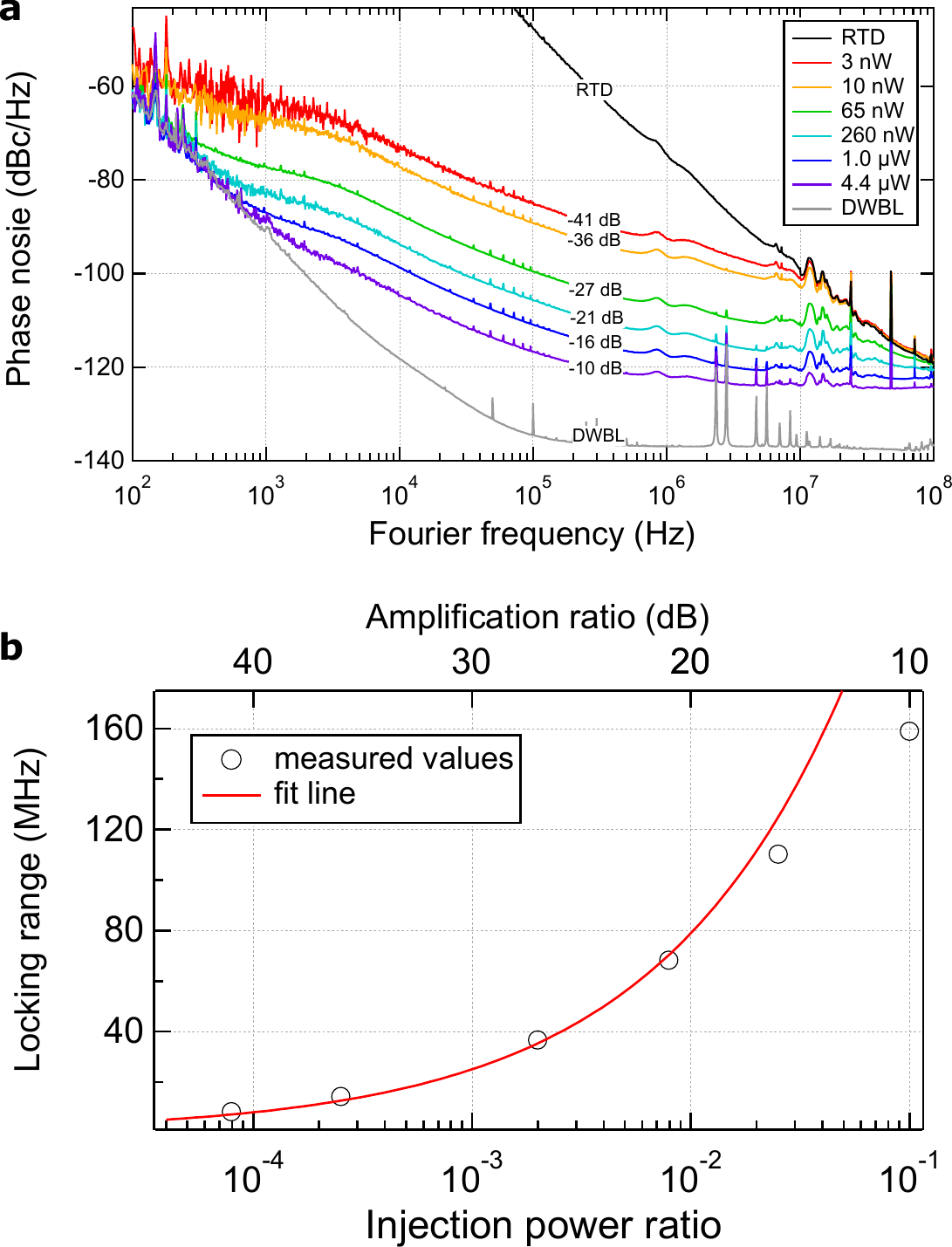}
    \caption{RTD injection locked phase noise and locking range versus power. \textbf{a} Injection locked phase noise of the waveguide RTD for different listed injection powers. The injection power ratio, relative to $P_\text{RTD}=40\,\upmu$W is overlayed on the corresponding phase noise trace. The RTD free running phase noise and DWBL phase noise are shown for reference. \textbf{b} Injection locking range at each injection power ratio determined by the intersection of the injected RTD phase noise with free running phase noise. The inverse of the power ratio is denoted on the top axis as an amplification ratio (gain). A fit line to Eq. \ref{lockrange}, excluding the top injection power ratio, is used to determine the RTD quality factor, $Q=165\pm5$. The error is the 67\% confidence interval of the fit parameter.}
    \label{fig_5}
\end{figure}

The low phase noise of the DWBL allowed the unambiguous measurement of the residual phase noise of injection locking. This is shown in Fig. \ref{fig_5}\textbf{a} for varying injection powers. 3\,nW of injected power was sufficient to synchronize the RTD to the DWBL, though under low-injection conditions, residual phase noise was much more significant. At 100\,Hz Fourier frequency, the phase noise of the RTD was reduced by nearly 90\,dB coinciding with the spectral purification described above. As the injection power was increased, the RTD was better able to follow the DWBL at higher Fourier frequencies, and the residual noise of injection decreased across all Fourier frequencies. Even at an injection power ratio of $-10$\,dB, however, measurable residual phase noise remained. The injected phase noise results approach the noise floor imposed by the mixer, $S_{\varphi}^\text{mixer}$, which was limited by large conversion loss, leading to a white noise floor of about -125\,dBc/Hz for 40\,$\upmu$W of incident RF power. The low-Fourier-frequency DWBL phase noise was the same as in Fig. \ref{fig_2}\textbf{a}, though under these experimental conditions, the white noise floor was lower.\footnote{The mixer used in this study had an unusually high conversion loss. Typical performance of this kind of mixer would limit the white noise floor to -145 dBc/Hz. Unfortunately, the mixer no longer meets its specifications.}

The point at which the residual phase noise crosses with the free running RTD phase noise was used to determine the injection locking range. Above this Fourier frequency, no synchronization between the oscillators occurs, and the RTD free running phase noise is preserved. The injection locking condition is defined by $|\sin{\theta}| \leq 1$ ($\theta$ is the injection phase offset). So the total injection locking range is given by,

\begin{equation}
    \Delta f_{lock} = f_\text{3dB}\rho.
    \label{lockrange}
\end{equation}

\noindent We used this relationship to experimentally determine $f_\text{3dB}=789$\,MHz; see Fig. \ref{fig_5}\textbf{b}. At higher injection power ratio the locking range was underestimated due to the mixer noise floor mentioned above, and further, these results were extrapolated, due to limited ESA bandwidth. Regardless, the lower power data are well described by Eq. \ref{lockrange} and allow us to characterize the RTD $Q$. The reciprocal of the injection power ratio can be thought of as an amplification ratio. Similarly, the locking range can be thought of as an amplifier bandwidth. The result is an injection-locking amplifier which trades gain for bandwidth. Over 40\,dB of amplification could be achieved with no active stabilization of the RTD bias. Larger gains can be realized by stabilizing the injection locking phase\cite{liu2020a,suzuki2022}.

\begin{figure}[t]
    \centering
    \includegraphics[width=\linewidth]{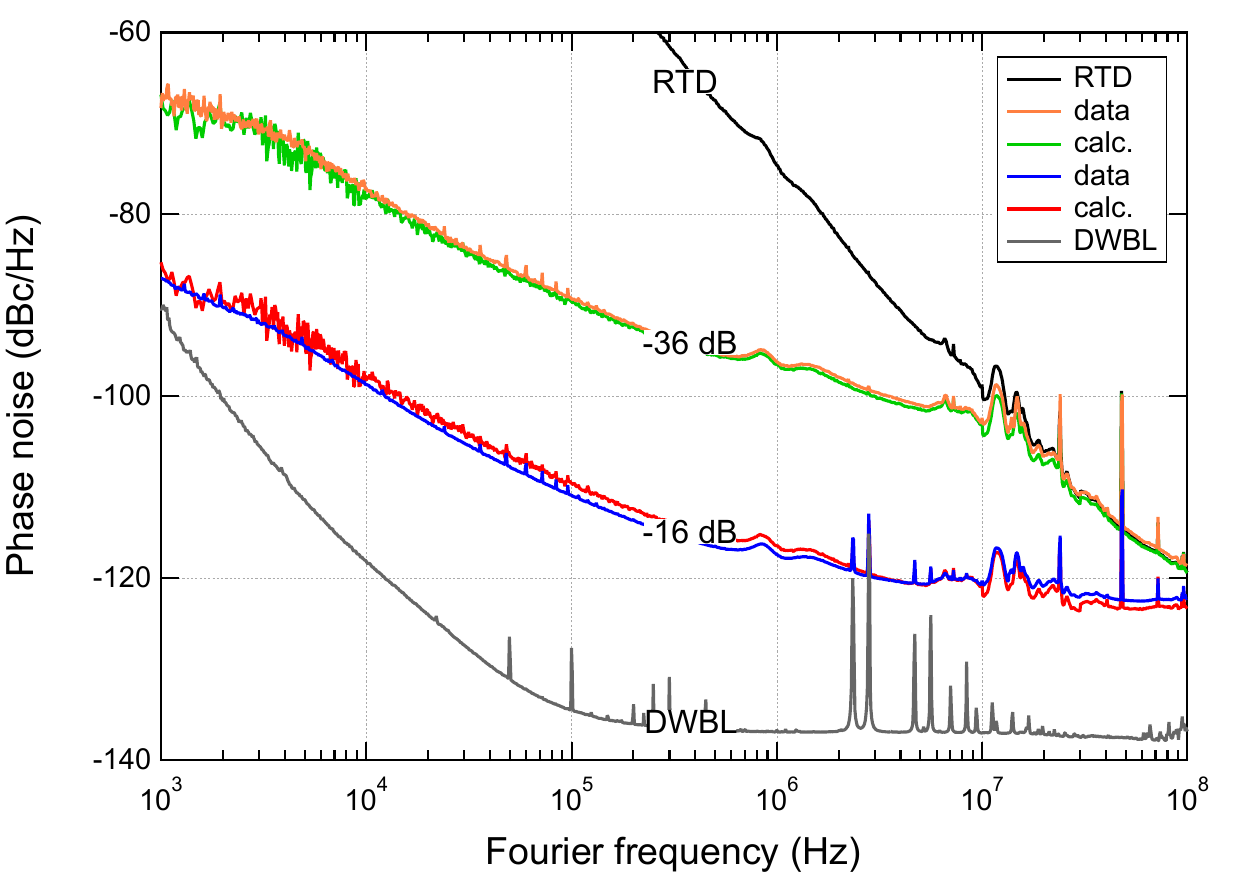}
    \caption{RTD residual injection noise data and calculations. Two injection noise traces from Fig. \ref{fig_5}\textbf{a} are reproduced here, labeled as data and by their injection power ratio. Evaluations of Eq. \ref{injmodel} for both power ratios are displayed to show good quantitative agreement between the model and data. Again, the free running RTD phase noise and DWBL phase noise are shown for reference.}
    \label{fig_6}
\end{figure}

Injection locking theory can be used to predict the phase noise of the synchronized RTD. The following relationship dictates the measured phase noise, 
\begin{equation}
    S_{\varphi}^\text{sync}(f) = \frac{\left( f/f_\text{3dB} \right)^2S_{\varphi}^\text{RTD}(f) + \rho^2\cos^2(\theta)S_{\varphi}^\text{DWBL}(f)}{\left( f/f_\text{3dB} \right)^2+\rho^2\cos^2(\theta)}
    \label{injmodel}
\end{equation}
\noindent which depends on the injection locking phase, $Q$, Fourier frequency, and free running phase noises of the RTD and DWBL. We used this relationship to calculate the expected phase noise of the injected RTD and found excellent agreement with the measured data. These comparisons are shown in Fig. \ref{fig_6}. We isolated two of the measured injection power ratios, 20\,dB apart, and compared the expected phase noise from evaluating Eq. \ref{injmodel}. Data and calculation agreed within 1\,dB across all Fourier frequencies and for all injection power ratios. The various spurs and features of the injected oscillator phase noise were inherited from both the DWBL and the RTD free running. To what degree the features were present depended on the injection power ratio and Fourier frequency.

The residual noise of injection can be calculated by removing the $S_{\varphi}^\text{DWBL}$ term in the numerator of Eq. \ref{injmodel}. Residual noise is the fundamental performance limitation of an injection locking amplifier. The RTD free running phase noise and $Q$ are the necessary quantities for evaluating an RTD based injection locking amplifier. As discussed in section \ref{sec_leeson}, the RTD free running phase noise can be deduced by combining electrical measurements of the RTD $b_{-1}/f$ and $b_{0}$ with Eq. \ref{leeson}. The quality factor, $Q$, can also be calculated from RTD equivalent circuit simulations \cite{asada2010}. All of these quantities taken together determine the residual noise of an RTD used as an injection amplifier without any direct terahertz phase noise measurements. Thus, the present experimental system is only required for evaluating the performance of a real injection locking amplifier. 

\section{Discussion}
At 260\,GHz, it is important to compare the performance of injection amplification with commercially available waveguide amplifiers. A commercial WR3.4 waveguide amplifier (VDI WR3.4AMP) provides up to 25dB gain with a noise figure of 9\,dB for most of the band (220-330\,GHz) and a saturation output power over 1\,mW. When operated below saturation, direct amplification of the photomixed DWBL yields a noise floor lower than the detection limit. However, injection amplification can provide significantly more gain in a single stage. The bandwidth and maximum output power of the injection amplifier are significantly lower.

Lower maximum output power and higher residual noise floor can be addressed by utilizing an RTD array. RTD arrays have shown coherent summation of radiated waves leading to much higher terahertz power \cite{koyama2022,tang2025,tang2025a}. Arrays also have a lower free running phase noise due to the benefits of mutual injection \cite{chang1997}. As a result, one can expect lower residual noise of injection, and higher output powers from RTD arrays used as injection amplifiers. The improved performance would come at the cost of narrower injection locking bandwidth.

A significant advantage of the RTD injection amplifier is its scalability to higher frequency. While the technologies underlying waveguide amplifiers can scale to higher frequency, it comes with performance degradation \cite{cheron2022,jyo2025}. RTD advancements to higher frequencies above 1\,THz however, seem very favorable \cite{suzuki2024}. 

We also point out the potential use of a narrow bandwidth, high gain injection locking amplifier in terahertz receivers. Tracking oscillators as well as tunable filters with gain can both be realized  by utilizing the same injection locking architecture presented here.

One other component in the RTD injection amplifier is critical to consider when thinking about scaling to higher frequency: the circulator. While it is likely that this passive component can be designed for higher bands, rectangular waveguide losses increase dramatically above 1\,THz. An integrated oscillator and injeciton amplifier design may be required to mitigate propagation loss. Circulators are also not strictly required to realize an injection amplifier. Back facet injection of RTDs has been demonstrated \cite{suzuki2024} and is an intriguing prospect for use at all frequencies.

Lastly, this study makes clear that increases in photomixed injection power enables lower phase noise. As significant improvements are made in the WR-3.4 band for photomixed power and amplifiers \cite{qian2024,cheron2022}, the phase noise of an injection amplified signal becomes more favorable. Furthermore, if limitations in photomixer power generation at high frequency persist, the higher low-frequency power can be exploited via the use of subharmonic injection locking \cite{asada2019}.

\section{Conclusion}
In summary, we have characterized the terahertz radiation of a waveguide RTD, including its free running phase noise. We analyzed the RTD phase noise and frequency fluctuations in detail characterizing the sources of noise finding the Leeson effect describes the observed behavior well. We then built and demonstrated an RTD injection locking amplifier using waveguide components. We measured the residual phase noise of the injection locking process and compared it to theoretical expectations.

We propose phase noise analysis of future RTD oscillators based purely on electrical characteristics of the diode. To our knowledge, this is the first such in depth analysis of RTD terahertz radiation phase noise and certainly the widest band (in terms of Fourier frequency) reported in the literature.

This study demonstrates over 40\,dB of amplification of a 260\,GHz wave for nanowatt-level input power. The combined photomixed and injection locked source provides a route to low phase noise and high power sources above 1\,THz. These sources are critical to the development of radioastronomy, imaging, molecular spectroscopy, radar, wireless communications, and other terahertz applications.

\section*{Acknowledgments}
We would like to thank Dr. Kazuisao Tsuruda and ROHM Semiconductor Inc. for use of the waveguide RTD central to this study. We also thank John Dorighi and Keysight for use of the SSA-X for phase noise measurements.
 


 





\end{document}